\documentclass[journal]{IEEEtran}

\usepackage[pdftex]{graphicx}
\usepackage{amsmath}
\usepackage{amssymb}
\usepackage{eqparbox}
\usepackage{algorithm}
\usepackage{algorithmic}
\usepackage{cite}
\usepackage{bm}
\usepackage{subfigure}
\usepackage{cases}
\usepackage{color,soul}

\begin{document}

\title{\LARGE Energy-Efficient Design for a NOMA assisted STAR-RIS Network with Deep Reinforcement Learning}
\author{Yi~Guo,~Fang~Fang,~\IEEEmembership{Memeber,~IEEE,}~Donghong~Cai,~\IEEEmembership{Memeber,~IEEE,}~and~Zhiguo~Ding,~\IEEEmembership{Fellow,~IEEE} \thanks{Yi Guo is with Xi’an Institute of Optics and Precision Mechanics of Chinese Academy of Sciences, Xi'an 710119, P.R.China (e-mail: guoyi@opt.ac.cn).}
\thanks{Fang Fang is with the Department of Engineering, Durham University,
Durham DH1 3LE, U.K. (e-mail: fang.fang@durham.ac.uk).}
\thanks{Donghong Cai is with the College of Cyber Security, Jinan University, Guangzhou 510632, P.R.China (e-mail: dhcai@jnu.edu.cn).}
\thanks{Zhiguo Ding is with the School of Electrical and Electronic Engineer-
ing, The University of Manchester, Manchester M13 9PL, U.K. (e-mail: zhiguo.ding@manchester.ac.uk).}
}
\maketitle


\begin{abstract}
Simultaneous transmitting and reflecting reconfigurable intelligent surfaces (STAR-RISs) has been considered as a promising auxiliary device to enhance the performance of the wireless network, where users located at the different sides of the surfaces can be simultaneously served by the transmitting and reflecting signals. In this paper, the energy efficiency (EE) maximization problem for a  non-orthogonal multiple access (NOMA) assisted STAR-RIS downlink network is investigated. Due to the fractional form of the EE, it is challenging to solve the EE maximization problem by the traditional convex optimization solutions. In this work, a deep deterministic policy gradient (DDPG)-based algorithm is proposed to maximize the EE by jointly optimizing the transmission beamforming vectors at the base station and the coefficients matrices at the STAR-RIS. Simulation results demonstrate that the proposed algorithm can effectively maximize the system EE considering the time-varying channels.
\end{abstract}

\begin{keywords}
NOMA-MISO, energy efficiency, DDPG, STAR-RISs.
\end{keywords}


\section{Introduction}
The meta-surfaces has been considered as one of key auxiliary devices in the future sixth generation (6G) wireless networks due to its substantial benefits, such as low-cost and low power consumption, communication coverage extension, and communication quality improvement \cite{2019arXiv190308925D}. With the development of corresponding fabrication technologies, two typical structures of meta-surfaces have been proposed recently\cite{2021arXiv210309104L}\cite{2021arXiv210109663X}, which are reconfigurable intelligent surfaces (RISs) and simultaneous transmitting and reflecting reconfigurable intelligent surfaces (STAR-RISs). Different from the RISs, which is commonly mentioned by its reflecting-only property, STAR-RISs can serve both sides of users located at its front and back, by simultaneously transmitting and reflecting the incident signals. Motivated by the attractive advantages of STAR-RISs, extensive research has been devoted to adopting STAR-RISs in exploiting the novel communication framework to achieve smart radio environments. 

Non-orthogonal multiple access (NOMA) is a promising 6G technology to achieve high spectrum efficiency and high energy efficiency\cite{9197675}\cite{8535085}. NOMA assisted STAR-RISs has been envisioned as a future promising wireless network structure.
As one of performance indicators for future wireless networks, the improvement of energy efficiency (EE) is important to avoid energy overhead and achieve green communications in 6G\cite{2021arXiv210101588M}. However, solving the EE maximization problem by finding the global optimal solution is challenging due to the fractional form of the objective function and non-convex constraints. While inspired by the successful application of deep reinforcement learning (DRL) in solving a variety of wireless communication problems\cite{2020arXiv200210072H}\cite{2021arXiv210406007D}, we design a Deep Deterministic Policy Gradient (DDPG)-based algorithm to maximize EE in a NOMA- multiple-input and single-output (MISO) assisted STAR-RIS downlink network. 
The proposed algorithm can effectively achieve the maximum system EE by considering various transmission power at the base station (BS) and different sizes of the STAR-RIS.

\begin{figure}[t] 
\centering
\includegraphics[width=3.45in]{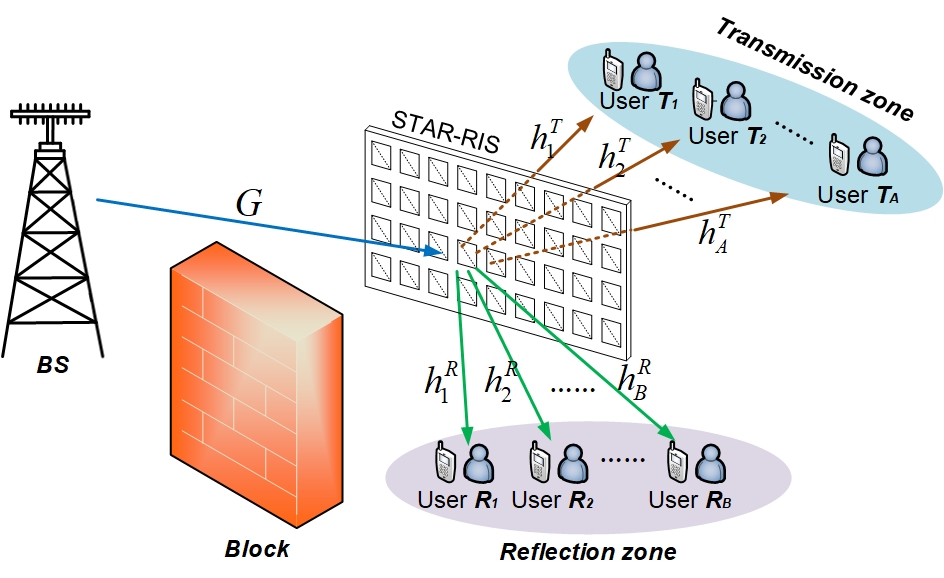}
\caption{Model of STAR-RIS assisted NOMA-MISO downlink network.}
\label{system model}
\end{figure}

\section{SYSTEM MODEL AND PROBLEM FORMULATION}
\subsection{System Model}
As shown in Fig. \ref{system model}, we consider a NOMA-MISO assisted STAR-RIS downlink network, where a BS with $\textit{M}$ antennas transmits the signals to multiple single-antenna users via a STAR-RIS which has $\textit{N}$ elements.
$T_{\Bar{a}}, \forall \Bar{a} \in\{1,2,\cdots,A\}$ denote the users located at the back of STAR-RIS in the transmission zone, while $R_{\Bar{b}}, \forall \Bar{b} \in\{1,2,\cdots,B\}$ are the users located in the reflection zone at the front of STAR-RIS, and where $\Bar{a}$ and $\Bar{b}$ are indices of the users in the transmission zone and reflection zone, respectively. We assume that the direct links between the BS and all the users are blocked by buildings or walls.

In this system, we assume that the STAR-RIS follows energy splitting (ES) protocol\cite{2021arXiv210109663X}\cite{2021arXiv210401421M}. The ES protocol indicates that every element at the STAR-RIS can simultaneously transmit and reflect the incident signals by adopting coefficients matrices at the same time. The coefficients matrices at the STAR-RIS contain amplitudes coefficients and phase shifts coefficients responding to the conditions of reflecting or transmitting signals. Under the ES protocol, the energy conservation law should be guaranteed \textemdash the sum of the energy of the transmitted and reflected signals must be equal to the incident signals' energy, which, ideally, defines the rule of amplitudes coefficients, i.e., $\beta_n^T + \beta_n^R = 1, \forall n \in\{1,2,\cdots,N\}$\cite{2021arXiv210109663X, 2021arXiv210309104L}, where $\beta_n^T$ and  $\beta_n^R$ denote the transmission and reflection amplitudes coefficients for $n$-th element at the STAR-RIS respectively. The coefficients matrices ${\bm{\Phi}}^{\tau} \in \mathbb{C}^{N \times N}$, $\tau \in \{T, R\}$ can be expressed as follows: 
\begin{equation}
\begin{split}
{\bm{\Phi}}^{\tau} = \text{diag}(\sqrt{{\beta}_1^{\tau}}e^{j{\theta}_1^{\tau}},\sqrt{{\beta}_2^{\tau}}e^{j{\theta}_2^{\tau}}, \cdots \sqrt{{\beta}_N^{\tau}}e^{j{\theta}_N^{\tau}}),
\end{split}
\end{equation}
where $\text{diag}(\cdot)$ presents the diagonal matrix. ${\theta}_n^{\tau}\in[0,2\pi)$, $\forall n \in\{1,2,\cdots,N\}$ denotes the phase shifts coefficient for the element at the STAR-RIS. 

In this paper, we consider that User $T_{\Bar{a}}$ and User $R_{\Bar{b}}$ are grouped together and served by the NOMA downlink transmission\cite{2021arXiv210309104L}. 
Define $\Omega = \{T_1,T_2,\cdots,T_A,R_1,R_2,\cdots,R_B\}$ as a user set by merging both users in the transmission zone and the reflection zone.
Let $y$ denote the superimposed signal transmitted from the BS to all the users, the received signal $y_{\epsilon}$ for User $\epsilon \in \Omega$ can be expressed as follows:
\begin{equation}\label{recieved signal}
y_{\epsilon} = \bm{h}_{\epsilon}^H \bm{\Phi}^{k(\epsilon)} \bm{G} \sum_{\upsilon \in \Omega} \bm{\omega}_{\upsilon} s_{\upsilon} + z, 
\end{equation}
where $\bm{h}_{\epsilon}^H \in \mathbb{C}^{1 \times N}$ denotes conjugate transpose of transmission or reflection channel between the STAR-RIS and User $\epsilon$. $\bm{G}\in \mathbb{C}^{N\times M}$ denotes the channel between the BS and the STAR-RIS, $n$ is the additive white Gaussian noise which follows $z \thicksim \mathcal{CN}(0, \sigma^2)$. $\bm{\omega}_{\upsilon} \in \mathbb{C}^{M \times 1}$ denotes the beamforming vector. $s_{\upsilon}$ is the signal symbol of User $\upsilon$ and we assume $\mathbb{E}(s_{\upsilon}) = 1$. $k(\cdot)$ is a function to get notation symbols which indicate transmission or reflection for coefficients matrices at the STAR-RIS:
\begin{subnumcases}{}
k(\epsilon) = T, & $if \ \epsilon = T_{\Bar{a}},$\\
k(\epsilon) = R, & $if \ \epsilon = R_{\Bar{b}}.$
\end{subnumcases}

For the NOMA transmission, successive interference cancellation (SIC) must be applied at the users. 
We assume the decoding order for all the users in $\Omega$ is $\chi = \{ U_{\Bar{c}}, \cdots, U_2, U_1 \}$, where $U_1, U_2, \cdots, U_{\Bar{c}} \in \Omega$ and $\Bar{c} = A + B$.
To apply SIC, the decoding order is assumed sequentially from the first element to the last element of $\chi$, i.e., from $U_{\Bar{c}}$ to $U_1$. Therefore, the achievable date rate at User $U_i \in \chi$ can be expressed as follows\cite{7555306}:
\begin{align}\label{user rate}
    R_{U_i} = \text{min}(R_{U_iU_i},R_{U_iU_j}),
\end{align}
where $R_{U_iU_i}$ denotes the decoding data rate at User $U_i$ when User $U_i$ decodes its own signal, and $R_{U_iU_j}$ denotes the decoding data rate at User $U_j \in \chi,j>i$ when User $U_j$ decodes User $U_i$'s signal. min($\cdot$) for data rate guarantees that SIC can be applied smoothly\cite{2021arXiv210603001Z}. $R_{U_iU_i}$ and $R_{U_iU_j}$ can be expressed as:
\begin{subnumcases}{}
R_{U_iU_i} = \text{log}_2(1+\frac{|\bm{h}_{U_i}^{H} \bm{\Phi}^{k(U_i)} \bm{G} \bm{\omega}_{U_i}|^2}{\sum\limits_{U_z \in \chi^{\prime}}|\bm{h}_{U_i}^{H} \bm{\Phi}^{k(U_i)} \bm{G} \bm{\omega}_{U_z}|^2 + \sigma^2}),\\ 
R_{U_iU_j} = \text{log}_2(1+\frac{|\bm{h}_{U_j}^{H} \bm{\Phi}^{k(U_j)} \bm{G} \bm{\omega}_{U_i}|^2}{\sum\limits_{U_z \in \chi^{\prime}}|\bm{h}_{U_j}^{H} \bm{\Phi}^{k(U_j)} \bm{G} \bm{\omega}_{U_z}|^2 + \sigma^2}),
\end{subnumcases}
where $\chi^{\prime} = \{U_{i-1},U_{i-2}, \cdots, U_j,\cdots,U_1\},i \leq \Bar{c}$ is a subset of $\Omega$, and the decoding order of users in $\{U_{\Bar{c}}, \cdots, U_{i+1}, U_{i}\}$ is priority than the decoding order of User $U_z$ in $\chi^{\prime}$.

\subsection{Problem Formulation}
In this paper, we aim to maximize the EE of the proposed downlink network.
The EE can be expressed as follows:
\begin{equation}\label{EE}
\eta_{EE} = \frac{B_w \sum\limits_{\epsilon \in \Omega}^{} R_\epsilon}{\frac{1}{\gamma} P_T+P_C},
\end{equation}
where $\gamma \in (0,1]$ denotes the efficiency of the power amplifier at the BS, and $P_C$ denotes total power consumption. $B_w$ denotes the transmission bandwidth. $P_T$ is the BS transmit power, which ideally can be expressed as the total power of all the users, i.e. $P_T = \sum\limits_{\epsilon \in \Omega}^{}||\bm{\omega}_{\epsilon}||^{2}$. 

Therefore, considering the related constrains for energy efficiency, the EE maximization problem can be formulated as:
\begin{subequations}\label{optimization}
\begin{align}
\label{problems}\mathop{max}\limits_{(\bm{\omega_\epsilon}, \bm{\Phi^{\tau}})} &\eta_{EE} \\
\label{power contrl}\text{s.t.} \quad &P_{T} \le P_{max}, \\
\label{beta contrl} &\beta^T_n + \beta^R_n = 1, \ \forall n \in\{1,2,\cdots,N\}, \\
\label{theta contrl}& 0\le \theta_{n}^{\tau} < 2\pi, \ \forall n \in\{1,2,\cdots,N\},\\
\label{target rate} & R_\epsilon \ge R_{min},
\end{align}
\end{subequations}
where constraint (\ref{power contrl}) describes that the transmission power limited at the BS, which indicates that the total power of the users can not exceed the maximum transmission power $P_{max}$ at the BS. Constraints (\ref{beta contrl}) and (\ref{theta contrl}) guarantee that amplitudes and phase shifts coefficients at the STAR-RIS will be adjusted within the reasonable ranges. Constraint (\ref{target rate}) guarantees that the data rate of all the users should meet the minimum data rate requirement of the system. 
Obviously, with the constrains and multiple variables, the EE maximizaiton problem  (\ref{optimization}) is non-convex, which is challenging to obtain the global optimal solution by using the traditional mathematical tools, such as convex optimization. 
To efficiently solve the problem (\ref{optimization}), we design a DRL-based algorithm to jointly optimize beamforming vectors at the BS and coefficients matrices including amplitudes and phase shifts at the STAR-RIS to maximize the EE. DRL is one of artificial intelligence (AI) technology, which can train fully autonomous agents via interacting with environment and applying specific optimal strategies, improving over time through trial and error\cite{DRLIntroduction}. With deep neural networks, DRL can solve more complex and high-dimensional optimal problems. As one of DRL, DDPG is applied to solve optimization problem in continuous space, which is suitable to solve our maximization problem.

\section{JOINT OPTIMAZATION WITH DDPG}
\subsection{Breif Introduction to DDPG}
Normally, there are four neural networks in DDPG: actor network, target actor network, critic network and target critic network, which two actor networks have the same parameters and structures, and the same features for both critic networks. A replay buffer is also used to store past experiences. One typical tuple of past experiences is organized as $(s^{(t)}, a^{(t)}, r^{(t)}, s^{(t+1)})$, where $s^{(t)}, a^{(t)}, r^{(t)}$ denotes state, action and reward in the current $t$-th training step, and $s^{(t+1)}$ is the state of the next step ($(t+1)$-th) obtained by executing action $a^{(t)}$ in the current environment. By randomly sampling $m_c$ tuples from the replay buffer, the parameters of the actor network can be updated by using the sampled policy gradient, and the critic network can be trained by minimizing the loss function\cite{8535085}\cite{2015arXiv150902971L}. A softly update method is adopted to update the parameters for both target networks. With four neural networks and their parameters update methods, DDPG model can constantly improve itself by repeating its backbone procedure\cite{2015arXiv150902971L}, to maximize the reward which can be specifically defined as the EE maximization in this work.

\subsection{Application of DDPG to EE optimization}
In this section, we briefly introduce the structures and process of our DDPG-based algorithm to the EE maximization. To apply DDPG to the maximization problem, the vectors both for action and state space, the reward function, the constraints normalization handling, and the algorithm process should be properly considered and designed in order to follow the DDPG operating rules.

According to the features in optimization problem (\ref{optimization}), we design the action vector, the state vector and the reward function as follows:
\begin{itemize}
  \item [1)] 
  Action vector:
  We select the beamforming vectors $\bm{\omega}_\epsilon^{(t)}$ and the coefficients matrices $\bm{\Phi}^{\tau,(t)}$ to define the action vector at the $t$-th training step.
  Note that $\bm{\omega}_\epsilon^{(t)}$ is a complex vector and the input vectors of neural networks should be real numbers. Thus we separately take the real part and imaginary part of $\bm{\omega}_\epsilon^{(t)}$ to construct one part of the action vector. Similarly, we take the real part and imaginary part of diagonal elements of $\bm{\Phi}^{\tau,(t)}$ to construct the rest part of the action vector. The action vector at the $t$-th training step can be presented as follows: 
  \begin{equation}\label{action space}
  \begin{split}
  & a^{(t)} =  \{ \text{Re}\{\bm{\omega}_\epsilon^{(t)}\}, \text{Im}\{\bm{\omega}_\epsilon^{(t)}\}, \text{Re}\{\bm{{\Phi}}^{\tau,(t)}_{n}\}, \text{Im}\{\bm{\Phi}^{\tau,(t)}_{n}\} \}, \\
  &  \forall\tau \in \{T,R\}, \ \forall\epsilon \in \Omega, \ \forall n \in \{1,2,\dots,N\},
  \end{split}
  \end{equation}
  where $\text{Re}\{\cdot\}$ and $\text{Im}\{\cdot\}$ present the real part and imaginary part of complex numbers respectively. $\bm{\Phi}^{\tau,(t)}_{n}$ denotes the $n$-th diagonal element of $\bm{\Phi}^{\tau,(t)}$.
  
  \item [2)]
  State vector:
  The state vector should fully present the status of the proposed communication system and consider the optimization problem  (\ref{optimization}). We design the state vector at the $t$-th traning step as follows:
  \begin{equation}\label{state space}
  \begin{split}
  &s^{(t)} = \{ R_\epsilon^{(t)}, ||\bm{\omega}_\epsilon^{(t)}||^2, |\bm{h}_\epsilon^{H,(t)} \bm{\Phi}^{k(\epsilon),(t)} \bm{G}^{(t)} |^2\},\\ &\forall\epsilon \in \Omega,
  \end{split}
  \end{equation}
  \item [3)]
  Reward function: Because our aim is to maximize the EE in (\ref{optimization}), it is naturally to use the EE as the reward for the $t$-th training step: $r^{(t)} = \eta_{EE}^{(t)}$.
\end{itemize}

\begin{algorithm}[t]
\caption{DDPG-based EE maximization}
\label{algorithm show}
\begin{algorithmic}[1]
\STATE Generate the actor network, the critic network, the target actor network and the target critic network with their parameters; 
\STATE Initialize the replay buffer $\mathcal{M}$ with the capacity C;
\FOR{episode $q = 1,2,...,E$}
\STATE Generate the channel $\bm{G}^{(q)}$ and $\bm{h}_{\epsilon}^{(q)},\forall\epsilon\in \Omega$ by (\ref{G channel}) and (\ref{h channel});
\STATE Initial $\bm{\omega}_{\epsilon}^{(1)}$,  $\bm{\Phi}^{\tau,(1)}$ and apply SIC to get $s^{(1)}$ by (\ref{state space});
\FOR{step $t = 1,2,...,S$}
\STATE Select the action $a^{(t)}$ from actor network based on the current state $s^{(t)}$; 
\STATE Explore $a^{(t)}$ by adding a random process $\mathcal{N}$;
\STATE Obtain $\bm{\hat{\omega}}_{\epsilon}^{(t)}$ by (\ref{normal w}), and obtain $\bm{\hat{\Phi}}^{\tau,(t)}$ by (\ref{normal phi});
\STATE Calculate the data rate $R_{\epsilon}^{(t)}$ at User $\epsilon$ by (\ref{user rate}); 
\IF {$R_{\epsilon}^{(t)} < R_{min}, \exists \epsilon \in \Omega$}
\STATE Calculate the reward $\Hat{r}^{(t)}$ with (\ref{EE}), (\ref{reward}) and (\ref{punishment2});
\ELSE 
\STATE Calculate the reward $\Hat{r}^{(t)}$ with (\ref{EE}), (\ref{reward}) and (\ref{punishment1});
\ENDIF
\STATE Construct a new state $s^{(t+1)}$ by (\ref{state space});
\STATE Store \{$s^{(t)}$, $a^{(t)}$, $\Hat{r}^{(t)}$, $s^{(t+1)}$\} to the replay buffer $\mathcal{M}$;
\STATE Randomly sample $m_c$ tuples from the replay buffer $\mathcal{M}$, and update the parameters of the critic network and the actor network; 
\STATE Softly update the parameters of the target actor network and the target critic network;
\STATE $s^{(t)} = s^{(t+1)}$
\ENDFOR
\ENDFOR
\end{algorithmic}
\end{algorithm}

To solve problem (\ref{optimization}), we propose a DDPG-based joint maximization algorithm shown in Algorithm \ref{algorithm show}. 
In this algorithm, each neural network is fully connected and sequentially comprises  the input layer, the hidden layer, the batch normalization layer, the hidden layer and the output layer. Regarding the actor neural network, the dimension of the input layer depends on the size of the state vector. In addition, the rectified linear activation (ReLU) function is used in the batch normalization layer, and the hyperbolic tangent (tanh) function is used in the second hidden layer. For the critic neural network, the state vector and the actor vector are fed to two individual hidden layers and two batch normalization layers, then the output of two batch normalization layers are concatenated together activated by the Relu function.
While the ReLU function is used in the second hidden layer for the critic neural network.
All the hidden layers contain 300 neurons in this paper. The learning rate for the critic network and the actor network are 0.002 and 0.001 respectively.

Considering that a build-in constraint structure in a neural network hardly meets the requirements of the constraints in (\ref{optimization}). It is possible to design a constraints handling process to normalize the beamforming vectors and the coefficients matrices in the original action vector which is the output of the actor neural network.

Note that the output range of the actor neural network is $(-1,1)$ due to the tanh activation function, the action vector $\bm{\omega}_\epsilon^{(t)}$ should be normalized in every training step to successfully apply the EE calculation and satisfy the constraints. Thus we normalize the beamforming vectors $\bm{\omega}_\epsilon^{(t)},\forall \epsilon \in \Omega$ as follows:  
\begin{align}
\label{normal w}  & \hat{\bm{\omega}}_\epsilon^{(t)} = \sqrt{\lambda_\epsilon^{(t)}} \bm{\omega}_\epsilon^{(t)},
\end{align}
where
\begin{subnumcases}{}
    \lambda_\epsilon^{(t)} = \frac{\hat{P}_\epsilon^{(t)}}{P_\epsilon^{(t)}},\\
    P_\epsilon^{(t)} = ||\bm{\omega}_\epsilon^{(t)}||^2,\\
    \hat{P}_\epsilon^{(t)} = \frac{||\bm{\omega}_\epsilon^{(t)}||^2}{(A+B)||\bm{\omega}_{tanh}^{max}||^2} \cdot P_{max},    \\
    |\text{Re}\{\bm{\omega}_{tanh}^{max}\}|,  |\text{Im}\{\bm{\omega}_{tanh}^{max}\}| \in \bm{1}^{M\times1},
\end{subnumcases}
where $||\bm{\omega}_{tanh}^{max}||^2$ presents the achievable maximum value for $||\bm{\omega}_\epsilon^{(t)}||^2$ in the range of the tanh function. $P_\epsilon^{(t)}$ denotes the transmission power of the beamforming vector $\omega_{\epsilon}^{(t)}$ organized by the action vector. $\hat{P}_\epsilon^{(t)}$ denotes the transmission power of the normalized beamforming vector $\hat{\omega}_{\epsilon}^{(t)}$. $\hat{P}_\epsilon^{(t)}$ is the ratio of $P_{max}$, which can further satisfy:
\begin{align}
\label{satisfed power} P_T^{(t)} = \sum\limits_{\epsilon \in \Omega}^{}\hat{P}_\epsilon^{(t)} \le P_{max},
\end{align}
where (\ref{satisfed power}) guarantees the constrains (\ref{power contrl}). Thus, Based on (\ref{normal w}), we can create a new normalized beamforming vector by the power ratio $\lambda_\epsilon^{(t)}$. Meanwhile, $\hat{\omega}_{\epsilon}^{(t)}$ maintains the same direction with $\omega_{\epsilon}^{(t)}$.

Similarly, the normalized coefficients matrices $\hat{\bm{\Phi}}^{\tau,(t)}$, $\forall \tau \in \{T, R\}$ as follows:
\begin{equation}\label{normal phi}
\begin{split}
& \hat{\bm{\Phi}}^{\tau,(t)} = \text{diag}(\sqrt{{\hat{\beta}}_1^{^{\tau,(t)}}}e^{j{\hat{\theta}}_1^{^{\tau,(t)}}},\sqrt{{\hat{\beta}}_2^{^{\tau,(t)}}}e^{j{\hat{\theta}}_2^{^{\tau,(t)}}}, \cdots,\\
&\sqrt{{\hat{\beta}}_N^{^{\tau,(t)}}}e^{j{\hat{\theta}}_N^{^{\tau,(t)}}}),
\end{split}
\end{equation}
where $\forall n \in\{1,2,\cdots,N\}$,
\begin{subnumcases}{}
\label{theta call}\hat{\theta}_n^{\tau,(t)} = \arctan(\frac{\text{Im}\{\bm{\Phi}^{\tau,(t)}_{n} \}}{\text{Re}\{\bm{\Phi}^{\tau,(t)}_{n} \}}), \\
\label{beta call}\hat{\beta}^{\tau,(t)}_{n} = \frac{|\bm{\Phi}^{\tau,(t)}_{n}|^2}{|\bm{\Phi}^{T,(t)}_n|^2 + |\bm{\Phi}^{R,(t)}_n|^2}, 
\end{subnumcases}
where (\ref{theta call}) guarantees that the polar form of $\bm{\Phi}^{\tau,(t)}_{n}$ maintain the same radians with its rectangular form. (\ref{beta call}) guarantees that the constrains (\ref{beta contrl}) can be satisfied as: $\hat{\beta}^{T,(t)}_{n} + \hat{\beta}^{R,(t)}_{n} = 1$.

Furthermore, a punishment rule is designed for reward at the $t$-th training step, which can be expressed as:
\begin{align}\label{reward}
    \hat{r}^{(t)} = \zeta \ r^{(t)},
\end{align}
where $\zeta$ is a punishment for the reward, which can be presented as:
\begin{subnumcases}{}
\zeta = 1,&$if \ R_{\epsilon}^{(t)} \geqslant R_{min}, \forall \epsilon \in \Omega $ \label{punishment1},\\
\zeta = -|R_{\epsilon,min}^{(t)}-R_{min}|,&$if \ R_{\epsilon}^{(t)} < R_{min}, \exists \epsilon \in \Omega$ \label{punishment2},
\end{subnumcases}
where $R_{\epsilon,min}^{(t)}$ presents the minimum data rate of all the users at the $t$-th training step. (\ref{punishment2}) punishes the reward in the negative way if any data rate of the users is less than the data rate requirement, which guarantees that the constraint (\ref{target rate}) can be satisfied. (\ref{punishment1}) denotes that the reward remains the value of the EE if all the users' data rate meet the data rate requirement. (\ref{punishment1}) and (\ref{punishment2}) both affect the reward in training. With (\ref{reward}), the DDPG model adjust the parameters to avoid the negative reward and try to achieve higher EE value through training.  

\section{NUMERICAL RESULTS}
In this section, we present the performance of the proposed joint maximization algorithm. Specifically, Based on the related works\cite{2021arXiv210401421M}, we model the channels gain $\bm{G}^{(q)}$ and $\bm{h}_{\epsilon}^{(q)},\forall\epsilon \in \Omega$ as Rician fading channel:
\begin{subequations}
\begin{align}
\label{G channel} & \bm{G}^{(q)} = \sqrt{\frac{\rho_0}{d_G^{\alpha_{BR}}}}(\sqrt{\frac{K_{BR}}{1+K_{BR}}} \bm{G}^{LoS} + \sqrt{\frac{1}{1+K_{BR}}} \bm{G}^{nLoS}), \\
\label{h channel} & \bm{h}_{\epsilon}^{(q)} = \sqrt{\frac{\rho_0}{d_{\epsilon}^{\alpha_{RU}}}}(\sqrt{\frac{K_{AU}}{1+K_{RU}}} \bm{h}_{\epsilon}^{LoS} + \sqrt{\frac{1}{1+K_{RU}}} \bm{h}_{\epsilon}^{nLoS}),
\end{align}
\end{subequations}
where $\rho_0$ denotes the path loss at a reference distance of 1 meter. $\alpha_{BR}$, $\alpha_{RU}$ are path loss exponents. $d_G$, $d_k$ denote distance between the STAR-RIS and the BS as well as distance between the STAR-RIS and the users respectively. $K_{BR},K_{RU}$ denote the Rician factors. $\bm{G}^{LoS}$ and $\bm{h}_{\epsilon}^{LoS}$ are the line-of-sight (Los) components, while $\bm{G}^{nLoS}$ and $\bm{h}_{\epsilon}^{nLoS}$ are the none-line-of-sight (nLos) components both following Rayleigh fading. It is worth to point out that $\bm{G}^{(q)}$ and $\bm{h}_{\epsilon}^{(q)}$ are generated at every episode in training to simulate varying channels.
Furthermore, the mainly setting parameters are demonstrated in Table \ref{table_p}.

\begin{table}[htb]
\begin{center}
\caption{SIMULATION PARAMETERS}
\label{table_p}
\begin{tabular}{|c|c||c|c|}
\hline  
parameter&value&parameter&value\\
\hline  
$d_G$&50 meters&$d_k$&(5,10) meters\\
\hline  
$\rho_0$&-30 dB& $\gamma$ & 0.35\\
\hline  
$P_c$&40 dBm& $\bm{G}^{LoS}, \bm{h}_k^{LoS}$ & 1\\
\hline  
$\alpha_{BR}, \alpha_{RU}$ & 2.2, 2.5 & $\sigma^2$ &-80 dBm\\
\hline 
$K_{BR},K_{RU}$&10& $B_w$ & 180 $k$Hz\\
\hline 
$C$ & 10000 & $m_c$ & 32\\
\hline 
\end{tabular}
\end{center}
\end{table}

\begin{figure}[t] 
\centering
\includegraphics[width=1\columnwidth]{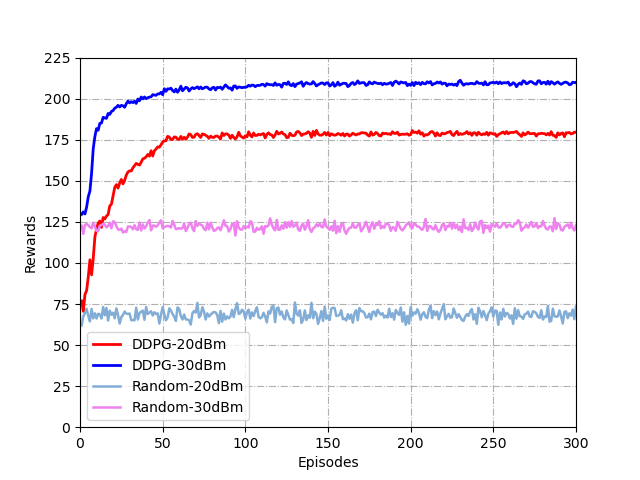}
\caption{Rewards versus Episodes with $M = 10$, $N = 30$, $A = B = 2$, $R_{min}$ = 0.1 bps/Hz, as well as different power at the BS}
\label{Rewards versus episode}
\end{figure}

Fig. \ref{Rewards versus episode} demonstrates the convergence of the proposed algorithm through the training episodes separately considering the time-varying channel with $P_{max} = 20$ dBm and $P_{max} = 30$ dBm at the BS. Each side of the  STAR-RIS has two users, and the minimum data rate requirement is set to 0.1 bps/Hz. From Fig. \ref{Rewards versus episode}, we can see that the rewards rise dramatically and then remain at a relatively high value with the increase of episodes for both transmission power. As one of the benchmarks in our simulation, a random coefficients scheme for the STAR-RIS remains poor performance with the episodes increases, which indicates that our proposed algorithm can significantly maximize the EE for the proposed downlink network.

\begin{figure}[t] 
\centering
\includegraphics[width=0.9\columnwidth]{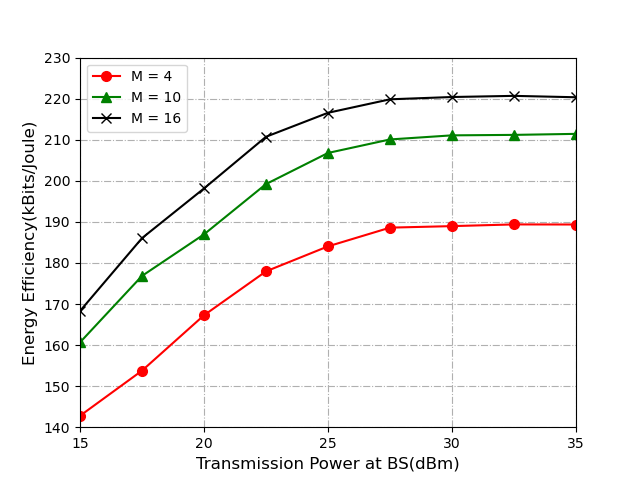}
\caption{EE versus transmission power at the BS with $R_{min}$ = 0.1 bps/Hz, $N = 30, A = B = 2$, as well as different antennas at the BS}
\label{EE versus power}
\end{figure}

Fig. \ref{EE versus power} shows EE versus the transmission power at the BS with the variable number of antennas at the BS. The number of elements at the STAR-RIS is 30. The user numbers and the data rate requirement are same with Fig. \ref{Rewards versus episode}. From Fig. \ref{EE versus power}, we can see that, as maximum transmitted power increases, EE increases to a peak value and remains, which indicates that EE can not grow continually with the constant growth of power at the BS. Moreover, the improvement of performance continuously gets smaller with the number of antennas increases. This is because the feasible domain of each channel between antennas get narrowed under the same power.

\begin{figure}[t] 
\centering
\includegraphics[width=1\columnwidth]{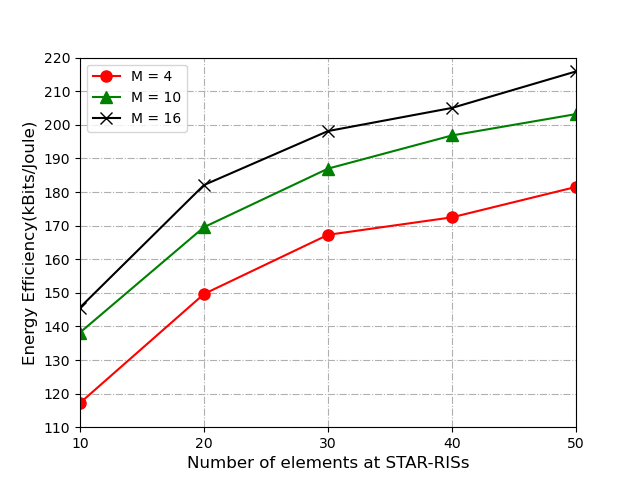}
\caption{EE versus elements at the STAR-RIS with $R_{min}$ = 0.1 bps/Hz, $A = B = 2$, as well as different antennas at the BS}
\label{EE versus elements}
\end{figure}

In Fig. \ref{EE versus elements}, we present the EE performance versus the number of the elements at the STAR-RIS with 20 dBm at the BS. It can be observed that the system EE increases with the number of the elements at the STAR-RIS. 

\section{Conclusion}
In this paper, we have studied a joint EE maximization problem for a NOMA-MISO assisted STAR-RIS downlink network. We have designed a DDPG-based algorithm to jointly optimize the beamforming vectors at the BS and the coefficients matrices at the STAR-RIS to maximize EE. The numerical results have validated the effectiveness and convergence of the proposed algorithm considering the time-varying channel.
Moreover, we have analyzed the trend of EE with different transmission power at the BS and various elements at the STAR-RIS.

\bibliographystyle{IEEEtran}
\bibliography{IEEEabrv,biblio_rectifier}

\begin{thebibliography}{10}
\providecommand{\url}[1]{#1}
\csname url@samestyle\endcsname
\providecommand{\newblock}{\relax}
\providecommand{\bibinfo}[2]{#2}
\providecommand{\BIBentrySTDinterwordspacing}{\spaceskip=0pt\relax}
\providecommand{\BIBentryALTinterwordstretchfactor}{4}
\providecommand{\BIBentryALTinterwordspacing}{\spaceskip=\fontdimen2\font plus
\BIBentryALTinterwordstretchfactor\fontdimen3\font minus
  \fontdimen4\font\relax}
\providecommand{\BIBforeignlanguage}[2]{{%
\expandafter\ifx\csname l@#1\endcsname\relax
\typeout{** WARNING: IEEEtran.bst: No hyphenation pattern has been}%
\typeout{** loaded for the language `#1'. Using the pattern for}%
\typeout{** the default language instead.}%
\else
\language=\csname l@#1\endcsname
\fi
#2}}
\providecommand{\BIBdecl}{\relax}
\BIBdecl

\bibitem{2019arXiv190308925D}
M.~{Di Renzo}, M.~{Debbah}, D.-T. {Phan-Huy}, A.~{Zappone}, M.-S. {Alouini},
  C.~{Yuen}, V.~{Sciancalepore}, G.~C. {Alexandropoulos}, J.~{Hoydis},
  H.~{Gacanin}, J.~{de Rosny}, A.~{Bounceu}, G.~{Lerosey}, and M.~{Fink},
  ``{Smart Radio Environments Empowered by AI Reconfigurable Meta-Surfaces: An
  Idea Whose Time Has Come},'' \emph{arXiv e-prints}, p. arXiv:1903.08925, Mar.
  2019.

\bibitem{2021arXiv210309104L}
Y.~{Liu}, X.~{Mu}, J.~{Xu}, R.~{Schober}, Y.~{Hao}, H.~V. {Poor}, and
  L.~{Hanzo}, ``{STAR: Simultaneous Transmission And Reflection for
  360{\textdegree} Coverage by Intelligent Surfaces},'' \emph{arXiv e-prints},
  p. arXiv:2103.09104, Mar. 2021.

\bibitem{2021arXiv210109663X}
J.~{Xu}, Y.~{Liu}, X.~{Mu}, and O.~A. {Dobre}, ``{STAR-RISs: Simultaneous
  Transmitting and Reflecting Reconfigurable Intelligent Surfaces},''
  \emph{arXiv e-prints}, p. arXiv:2101.09663, Jan. 2021.

\bibitem{9197675}
F.~Fang, Y.~Xu, Q.-V. Pham, and Z.~Ding, ``Energy-efficient design of irs-noma
  networks,'' \emph{IEEE Trans. Veh. Technol.}, vol.~69, no.~11, pp.
  14\,088--14\,092, 2020.

\bibitem{8535085}
K.~Yang, N.~Yang, N.~Ye, M.~Jia, Z.~Gao, and R.~Fan, ``Non-orthogonal multiple
  access: Achieving sustainable future radio access,'' \emph{IEEE Commun.
  Mag.}, vol.~57, no.~2, pp. 116--121, 2019.

\bibitem{2021arXiv210101588M}
B.~{Mao}, F.~{Tang}, K.~{Yuichi}, and N.~{Kato}, ``{AI based Service Management
  for 6G Green Communications},'' \emph{arXiv e-prints}, p. arXiv:2101.01588,
  Jan. 2021.

\bibitem{2020arXiv200210072H}
C.~{Huang}, R.~{Mo}, and C.~{Yuen}, ``{Reconfigurable Intelligent Surface
  Assisted Multiuser MISO Systems Exploiting Deep Reinforcement Learning},''
  \emph{arXiv e-prints}, p. arXiv:2002.10072, Feb. 2020.

\bibitem{2021arXiv210406007D}
Z.~{Ding}, R.~{Schober}, and H.~V. {Poor}, ``{No-Pain No-Gain: DRL Assisted
  Optimization in Energy-Constrained CR-NOMA Networks},'' \emph{arXiv
  e-prints}, p. arXiv:2104.06007, Apr. 2021.

\bibitem{2021arXiv210401421M}
X.~{Mu}, Y.~{Liu}, L.~{Guo}, J.~{Lin}, and R.~{Schober}, ``{Simultaneously
  Transmitting And Reflecting (STAR) RIS Aided Wireless Communications},''
  \emph{arXiv e-prints}, p. arXiv:2104.01421, Apr. 2021.

\bibitem{7555306}
Z.~Chen, Z.~Ding, X.~Dai, and G.~K. Karagiannidis, ``On the application of
  quasi-degradation to miso-noma downlink,'' \emph{IEEE Trans. Signal
  Process.}, vol.~64, no.~23, pp. 6174--6189, 2016.

\bibitem{2021arXiv210603001Z}
J.~{Zuo}, Y.~{Liu}, Z.~{Ding}, L.~{Song}, and H.~V. {Poor}, ``{Joint Design for
  Simultaneously Transmitting And Reflecting (STAR) RIS Assisted NOMA
  Systems},'' \emph{arXiv e-prints}, p. arXiv:2106.03001, Jun. 2021.

\bibitem{DRLIntroduction}
K.~Arulkumaran, M.~P. Deisenroth, M.~Brundage, and A.~A. Bharath, ``Deep
  reinforcement learning: A brief survey,'' \emph{IEEE Signal Process. Mag.},
  vol.~34, no.~6, pp. 26--38, 2017.

\bibitem{2015arXiv150902971L}
T.~P. {Lillicrap}, J.~J. {Hunt}, A.~{Pritzel}, N.~{Heess}, T.~{Erez},
  Y.~{Tassa}, D.~{Silver}, and D.~{Wierstra}, ``{Continuous control with deep
  reinforcement learning},'' \emph{arXiv e-prints}, p. arXiv:1509.02971, Sep.
  2015.

\end{thebibliography}

\end{document}